\documentclass[12pt]{article}

\newcommand{\fr}{\frac}
\newcommand{\lb}{\label}
\newcommand{\ti}{\tilde}
\newcommand{\be}{\begin{equation}}
\newcommand{\ee}{\end{equation}}
\newcommand{\ba}{\begin{array}}
\newcommand{\ea}{\end{array}}
\newcommand{\beqa}{\begin{eqnarray}}
\newcommand{\al}{\alpha}

\newcommand{\la}{\lambda}

\newcommand{\si}{\sigma}
\newcommand{\te}{\theta}
\newcommand{\del}{\partial}
\newcommand{\eeqa}{\end{eqnarray}}
\newcommand{\ep}{\epsilon}

\newcommand{\kd}{\delta}

\begin{document}
\title{}
\author{}
\date{}
\begin{flushright}
hep-th/0208043
\end{flushright}

\vspace{1cm}

\noindent
{\Large \bf Hamiltonian formulation of Noncommutative D3--Brane}

\vspace{1cm}
\noindent
{\footnotesize  \"{O}mer F. DAYI$^{a,b,}$}\footnote{E-mail 
addresses: dayi@gursey.gov.tr and  dayi@itu.edu.tr.}
{\footnotesize and Bar\i\c{s} YAPI\c{S}KAN $^{a,}$}\footnote{E-mail address: 
yapiska1@itu.edu.tr}

\vspace{10pt}

\noindent
$a)${\footnotesize \it Physics Department, Faculty of Science and
Letters, Istanbul Technical University,}

\noindent
\hspace{3mm}{\footnotesize \it  80626 Maslak--Istanbul,
Turkey.}

\vspace{10pt}

\noindent
$b)${\footnotesize \it Feza G\"{u}rsey Institute,}

\noindent
\hspace{3mm}{\footnotesize \it P.O.Box 6, 81220
\c{C}engelk\"{o}y--Istanbul, Turkey. }

\vspace{2cm}
\noindent
{\footnotesize 

\noindent
{\bf Abstract} 

\noindent
Lagrangians of the Abelian Gauge Theory and
 its dual are related in terms of a shifted
action. We show that in $d=4$ 
constrained Hamiltonian formulation of
the shifted action yields
Hamiltonian description of the dual theory,
without referring to  its Lagrangian.
We apply this method,
at the first order in the noncommutativity parameter $\te ,$
to the noncommutative $U(1)$
gauge theory possessing spatial noncommutativity.
Its dual theory is effectively  
a space--time noncommutative $U(1)$
gauge theory. However, we obtain a 
Hamiltonian formulation where time is commuting.
Space--time
noncommutative D3--brane worldvolume Hamiltonian
is derived as the dual of space noncommutative $U(1)$ gauge theory.
We show that a BPS like bound can be obtained and
it is saturated for configurations which 
are the same with the ordinary D3-brane 
BIon and dyon solutions.
  }
\newpage

\section{Introduction}

Noncommutative and ordinary gauge theory descriptions
of D--branes in a constant background B--field
are equivalent perturbatively in the noncommutativity 
parameter $\te^{\mu \nu}$\cite{sw}. To find evidence that this
equivalence is valid even nonperturbatively,
noncommutative D3--brane BIon and dyon
are studied in  \cite{mat}. Noncommutative D3--brane 
BIon configuration is attained
when open
string metric satisfies $G_{MN}={\rm diag} (-1,1,\cdots,1)$ where
$M,N=0,1,\cdots,9.$ This geometry is accomplished
allowing a background 
B--field on D3--brane worldvolume,
producing a noncommutativity parameter $\theta^{01}\neq 
0$ and  $\theta^{02}=\theta^{03}={\theta}^{ij}=0,$ 
where $i,j=1,2,3.$ 
At the lowest order in the string slope parameter
$\alpha^\prime$ 
and for slowly varying fields noncommutative
D3--brane is described
 in terms of noncommutative $U(1)$ 
gauge theory. Non--vanishing ${\theta}^{01}$ leads to noncommutative 
time in terms of Moyal bracket.
Thus the ordinary Hamiltonian formalism of this
system is obscure. However, 
owing to the fact that the theory is invariant 
under translations, an energy density
 is derived from Lagrangian which is utilized to 
write a Bogomol'nyi--Prasad--Sommerfeld (BPS) 
like bound.

When time is noncommuting with the spatial 
coordinates the usual Hamiltonian methods 
are not applicable.
To overcome this difficulty in \cite{gkm}
a new method is developed introducing
a  spurious 
time like variable. In this approach the energy 
is the same as the  one
derived from Lagrangian path integral formalism
of the original theory.

We would like to examine if 
one can find a Hamiltonian
in an ordinary phase space
for space--time noncommutative
D3--brane. It would yield  a
well defined energy.
This is possible if the Lagrangian with
noncommutative time can be considered as an object derived 
from an original theory whose time variable is 
commuting. Indeed, in string theories
noncommutative time parameter usually
appears in the actions
which are (S) dual of initial theories
with  commuting time\cite{sd-nc}.
Similarly, in \cite{grs}
noncomutative $U(1)$ gauge theory with
the noncommutativity parameter 
$\tilde{\theta}^{ij}=0,$ 
$\tilde{\theta}^{0i}\neq 0,$  is established as the   
dual theory of the one whose
noncommutativity parameter satisfies
$\theta^{ij}\neq 0,$ $\theta^{0i}= 0.$
 
Legendre transforming the Abelian gauge theory Lagrangian
in terms of dual gauge field and performing path
integration of the shifted action over the field strength lead to 
the Lagrangian formalism of the dual theory.
We will show that constrained Hamiltonian structure\cite{dirac}
 of the shifted
Abelian gauge theory action in $d=4$ provides us
Hamiltonian formalism of  the dual theory 
without referring to its Lagrangian. 
This method of bypassing Lagrangian of the dual
theory to derive its phase space formalism
is interesting in itself. Moreover, 
 it may be very useful
to treat the theories whose Lagrangian formalism is given
in terms of (effectively) noncommuting time variable.
In fact,
we apply it to
noncommutative $U(1)$ gauge theory with spatial 
noncommutativity and obtain phase space formulation
of dual gauge theory whose time coordinate is 
noncommuting in terms of Moyal bracket. 
On the other hand the dual Lagrangian is known
and it is originated  from a theory with a commuting time variable.
Thus, one can treat time variable as commuting, although effectively
time and spatial coordinates satisfy a non--vanishing 
Moyal bracket.
Hamiltonian and phase space
constraints of the both approaches coincide.
We deal only with
the first order approximation in noncommutativity 
parameter $\te^{\mu \nu}.$

Once Hamiltonian formulation of the dual theory
is achieved,  noncommutative D3--brane worldvolume 
 Hamiltonian formalism 
in static gauge can be written directly.
We define $\te$ dependent (noncommutative)
fields in terms of the usual phase space fields
and obtain a BPS
like bound on energy.
Saturation of this bound leads to equations in terms 
of $\te$ dependent fields. These are  solved when 
commutative fields satisfy ordinary BPS equations. 
This is similar to the observation that linearized and full 
Dirac--Born--Infeld (DBI) theories possess the 
same BPS states\cite{cm} with the same energy. 
However, in our case energies differ.

In Section 2 we study
 Abelian Gauge Theory in $d=4.$ 
We show
that the appropriately shifted Lagrangian can be studied as a constrained 
Hamiltonian system and reduced phase space method leads to Hamiltonian 
formulation of the 
dual theory. Thus, without referring to the Lagrangian we can obtain 
the Hamiltonian formalism of the dual theory.

In Section 3 
the method illustrated in Section 2
is utilized 
to derive  Hamiltonian formulation of 
the noncommutative  $U(1)$ gauge theory 
whose time coordinate is noncommuting with the spatial ones
in terms of Moyal bracket. We treat the gauge theory
with spatial noncommutativity as the original one.
Considering time as commuting in the dual theory
a phase space formulation
is found 
in terms of the usual methods.
These two Hamiltonian descriptions coincide. 

In Section 4  we  derive the
 Hamiltonian for D3--brane worldvolume
with a scalar field.
We introduce $\te$ dependent (noncommutative)
fields to put
this Hamiltonian in a
form suitable to derive a 
BPS like bound on energy.
Conditions to  saturate this bound are discussed.
It is shown that these conditions are solved
when commuting  fields 
are taken as D3--brane BIon or  dyon solutions.
For some specific choices of fields,
charges 
taking part in BPS bound are shown to be topological,
although $\te$ dependent.

The results obtained are discussed in the last section.

\section{Abelian Gauge Theory}

Abelian  gauge theory action in $d=4$ with the Minkowski metric
$g_{\mu\nu}={\rm diag}(-1,1,1,1),$ is
\be
\lb{so}
S_o=-\fr{1}{4g^2}\int d^4 x F_{\mu\nu}F^{\mu\nu},
\ee
where $F=dA.$
To implement duality
we introduce the dual gauge
field $A_D$ and  deal with the shifted action  
\be
\lb{lt1}
S_m=\int d^4 x (\fr{1}{4g^2}F_{\mu\nu} F^{\mu\nu}-\fr{1}{2}
\ep_{\mu\nu\rho\sigma} 
\del^\mu A_D^\nu  F^{\rho\sigma}).
\ee
Hereö we
treat $F$ as an independent  variable without  
requiring any relation with the 
gauge field $A.$  Performing path integral over $F$, which is equivalent
to solve the equations of motion for $F$ in terms of $A_D$ and replace it
in the action, leads to the dual action
\be
\lb{sd}
S_D=-\fr{g^2}{4}\int d^4 x {F_D}^{\mu\nu}F_{D\mu\nu}
\ee 
where $ F_D=dA_D $.
Constraint Hamiltonian structures
resulting from  
$S_o$ and $S_D$ are related by
canonical transformations\cite{loz}. 

We would like to study canonical formulation
of $S_m$ to demonstrate that Hamiltonian and constraints
of the dual action $S_D$ result directly.
Definition of canonical momenta
\be
P_{\mu\nu}=\fr{\kd S_m}{\kd (\del_0 F^{\mu\nu})};\
P_{D\mu}=\fr{\kd S_m}{\kd (\del_0 A_D^\mu )},
\ee
leads to the primary constraints,
\beqa
P_{\mu\nu}\approx 0, \lb{p11} \\
P_{D0}\approx 0 \lb{p21}, \\
\phi_i\equiv P_{Di}+\frac{1}{2}\ep_{ijk}F^{jk} \approx 0, \lb{p31}
\eeqa  
where $i,j,k=1,2,3,$ and
`$\approx $' denotes that constraints are  weakly vanishing.
The related canonical Hamiltonian is
\be
\lb{cham}
H_{mc}=\int 
d^3x\ [\fr{1}{2g^2}F^{0i}F_{0i}+\fr{1}{4g^2}F^{ij}F_{ij}-\fr{1}{2}
\ep_{ijk}\del^i{{A_D}^0}F^{jk}+\ep_{ijk}\del^i{A_D}^jF^{0k}].
\ee
By adding the primary constraints (\ref{p11})--(\ref{p31})
to the canonical Hamiltonian,
 in terms of some Lagrange multipliers 
$\alpha_i,\ \beta,\ \la_{ij},\ \kappa_i,$ one obtains the extended 
Hamiltonian\cite{dirac}
\be
H_{me} =H_{mc} +
\int d^3x\
[\al_iP_{0i}+\beta P_{D0}+\la_{ij}P_{ij}+\kappa_i \phi_i].
\ee

Denote that we use the definition
$$
\fr{\del{F_{\mu\nu}}}{\del{F_{\rho\si}}}={\delta_\mu}^\rho 
{\delta_\nu}^\si-
{\delta_\mu}^\si {\delta_\nu}^\rho,
$$
to calculate 
Poisson brackets of the primary 
constraints (\ref{p11})--(\ref{p31}) with $H_{me},$ which lead to the 
secondary constraints
\be
\lb{s11}
\ep_{ijk}\del^iF^{jk}\approx 0,
\ee
\be
\lb{s21}
\fr{1}{g^2}F_{0i}+\ep_{ijk}\del^j{A_D}^k \approx 0.
\ee
Constraints terminate here.
The constraint 
(\ref{p21})
is first class and the rest
(\ref{p11}), (\ref{p31}), (\ref{s11}), (\ref{s21}) 
are  second class. In the 
reduced phase space, obtained by setting 
all the second class constraints equal to 
zero strongly
and solving $F,\ P$ in terms of $F_D\ P_D$, 
the canonical Hamiltonian (\ref{cham}) becomes
\be
H_{D}=\int d^3x[\fr{1}{2g^2}P_{Di}P_{Di}+\fr{g^2}{4}F_{Dij}F_{Dij}].
\ee
Moreover, there are  the first class constraints
\be
P_{D0}\approx 0,\  \del_i  P_{Di}\approx 0.
\ee
Obviously this is the same with
the constrained Hamiltonian formalism 
of the dual theory (\ref{sd}).
Therefore we demonstrated that  one can obtain 
constrained Hamiltonian formulation of the
dual theory beginning from 
the shifted action (\ref{lt1}) bypassing the dual Lagrangian (\ref{sd}).

\section{Noncommutative $U(1)$ Gauge Theory}

Noncommuting variables should be treated as operators. However,
one can retain them commuting under the usual product
and introduce 
noncommutativity in terms of
the star product
\be
\label{star}
\ast\equiv\exp \frac{i\te^{\mu\nu}}{2} \Big(
{\stackrel\leftarrow\del}_{\mu} {\stackrel\rightarrow\del}_{\nu}
-{\stackrel\leftarrow\del}_{\nu}{\stackrel\rightarrow\del}_{\mu}
\Big),
\ee  
where $\te^{\mu \nu}$ is a constant parameter.
Now, the coordinates $x^\mu$ satisfy the Moyal bracket
\be
\lb{nc}
x^\mu\ast x^\nu -
x^\nu\ast x^\mu =\te^{\mu \nu}.
\ee

Noncommutative $U(1)$ gauge theory is given by the action
\be
\lb{gt}
S_{nc}=-\fr{1}{4g^2}\int d^4x\widehat{F}_{\mu\nu}
\ast \widehat{F}^{\mu\nu}
\ee
where we defined
\be
\lb{fn}
\widehat{F}_{\mu\nu}=\del_\mu\widehat{A}_\nu-\del_\nu\widehat{A}_\mu
-i\widehat{A}_\mu\ast\widehat{A}_\nu+i\widehat{A}_\nu \ast\widehat{A}_\mu .
\ee

Seiberg and Witten\cite{sw} showed that noncommutative 
gauge fields $\widehat{A}_\mu ,$ noncommutative gauge parameter
$\widehat{\la}$ and  the commuting ones $A_\mu ,\ \la$
are related under the  gauge transformations 
$\widehat{\kd},\ \kd$ as
\be
\lb{swc}
\widehat{A}(A)
+\widehat{\kd}_{\widehat{\la}}
\widehat{A}(A)=\widehat{A}(A+\kd_\la A) .
\ee
At the first order in $\te^{\mu\nu}$ it is solved to yield 
\be
\lb{sw}
\widehat{F}_{\mu \nu}=F_{\mu \nu}+
\te^{\rho \sigma} F_{\mu \rho}F_{\nu \si } 
 - \te^{\rho \sigma} A_\rho
\del_\si F_{\mu \nu }.
\ee
Thus the action (\ref{gt}) can be written at the first order in $\te^{\mu\nu}$
as
\be
\lb{or}
S_{nc}=-\fr{1}{4g^2}\int d^4x 
(F_{\mu\nu}F^{\mu\nu}+2\te^{\mu\nu}F_{\nu\rho}F^{\rho\si}F_{\si\mu}
-\fr{1}{2}\te^{\mu\nu}F_{\nu\mu}F_{\rho\si}F^{\si\rho}).
\ee
To implement duality transformation 
deal with the shifted action\cite{grs}
\be
\lb{as}
S=-S_{nc}+ \fr{1}{2}\int d^4x\ {A_D}^\mu\ep_{\mu\nu\rho\si}\del^\nu F^{\rho\si}
\ee
where F and $A_D$ are taken as independent field variables.

As in the commutative case dual action can be 
found by solving the field equations
for $F$  in 
terms of $F_D=dA_D$ 
and plugging it in the action (\ref{as}) 
leading to   
\be
\lb{od}
\tilde S =-\fr{g^2}{4}\int d^4x({F_D}^{\mu\nu}F_{D\mu\nu}
+2{\tilde\te}^{\mu\nu}F_{D\nu\rho}
F^{\rho\si}_D F_{D\si\mu}-\fr{1}{2}{\tilde\te}^{\mu\nu}
F_{D\nu\mu}F_{D\rho\si}F_D^{\si\rho})
\ee
where  ${\tilde \te}^{\mu\nu}={\ep}^{\mu\nu\rho\si}\te_{\rho\si}$.
At the first order in $\ti{\te}$ it can be derived from
the action
\be
\lb{gtt}
\ti{S}=-\fr{g^2}{4}\int d^4x\widehat{F}_{D\mu\nu}
\ti{\ast} \widehat{F}_D^{\mu\nu},
\ee
where $\ti{\ast}$ is given by (\ref{star}) by replacing
$\te$ with $\ti{\te}.$
For
$\te^{0i}=0$ and $\te^{ij}\neq 0$
the dual theory is  a 
gauge theory whose time variable 
is noncommuting in terms of the  Moyal bracket (\ref{nc})
with $\ti{\ast},$
because   $\tilde{\te}^{oi}\neq 0,\ \ti{\te}^{ij}=0.$ 
For a noncommuting 
time canonical formalism is obscure.
Thus we would like to bypass the 
dual action (\ref{od}) to obtain  a phase space 
formulation 
of the dual theory using the method illustrated
in Section 2.

Let $\te^{ij}\neq0$ and $\te^{i0}=0$
in the action (\ref{as}).
Definition of canonical momenta
\be
P_{\mu\nu}=\fr{\delta S}{\delta (\del_0F^{\mu\nu})}
\ee
\be
P_{D\mu}=\fr{\delta S}{\delta (\del_0{A_D}^\mu)}
\ee
leads to the primary constraints
\be
\lb{con0}
P_{\mu\nu}  \approx 0,
\ee
\be
\lb{con1}
P_{D0} \approx 0,
\ee
\be
\lb{con2}
P_{Di}+\fr{1}{2}\epsilon_{ijk}F_{jk} \approx 0
\ee
and the canonical Hamiltonian
\beqa
H_c&=&\int d^3x[-\fr{1}{2}\epsilon_{ijk}\del^i{A_D}^0F^{jk}+
\epsilon_{ijk}
\del^i{A_D}^jF^{0k}+\fr{1}{2g^2}F_{0i}F^{0i} \nonumber \\
&&+\fr{1}{4g^2}F_{ij}F^{ij}+\fr{1}{g^2}F^{0i}F_{ij}\te^{jk}F_{k0}+\fr{1}{2g^2}
F^{ij}F_{jk}\te^{kl}F_{li} \nonumber \\
&&-\fr{1}{4g^2}\te^{ij}F_{ji}F_{0l}F^{l0}
-\fr{1}{8g^2}\te^{ij}F_{ji}F_{kl}F^{lk}] \lb{uf} .
\eeqa
Preserving the primary constraints 
(\ref{con0})--(\ref{con2})
in time leads to secondary constraints
\be
\lb{con3}
\epsilon_{ijk}\del^iF^{jk} \approx 0
\ee
and 
\beqa
\Psi_i&\equiv &F^{0i}-F_{ij}\te^{jk}F_{k0}-
F^{0j}F_{jk}\te^{ki}\nonumber \\
&&-{1\over 2}\te^{jk}F_{kj}F_{0i}- g^2\epsilon_{ijk}
\del^j{A_D}^k \approx 0 \lb{con4},
\eeqa
which do not yield new constraints.
One can check that (\ref{con1}) is first class and
 the other constraints 
(\ref{con0}),
(\ref{con2}),
(\ref{con3}),
(\ref{con4})
 are second class.
In the reduced phase space where  second class 
constraints strongly vanish, the canonical Hamiltonian 
(\ref{uf}) becomes
\beqa
H_{nD}&=&\int 
d^3x[\fr{g^2}{4}F_{Dij}^2+\fr{1}{2g^2}P_{Di}^2-\fr{1}{2g^2}\ep_{ijk}
\te^{ij}P_D^kP_{Dl}^2 \nonumber \\
&& -\fr{g^2}{4}\ep_{ijk}\te^{ij}P_D^kF_{Dlm}^2-g^2
F_{Dij}P_D^j\te^{ik}\ep_{klm}F_D^{lm}], \lb{hnd}
\eeqa
if we solve $F,\ P$ in terms of $F_D$ and $P_D$.
Moreover, there are still 
the constraints 
\be
\lb{chd}
\del_iP_{Di}=0,\ P_{D0}=0,
\ee
which are  first class.
This Hamiltonian can be written in terms 
of $\tilde \te^{0i}=\ep^{ijk}\te_{jk}$ as 
\beqa
H_{nD}&=&\int 
d^3x[\fr{g^2}{4}F_{Dij}^2+\fr{1}{2g^2}P_{Di}^2-\fr{1}{2g^2}\tilde\te_{0i}
P_D^iP_{Dj}^2\nonumber \\ 
&&-\fr{g^2}{4}\tilde\te_{0i}P_D^iF_{Djk}^2
+g^2\tilde
\te^{0i}F_{Dji}F_{Djk}P_D^k] \lb{sla}
\eeqa

On the other hand,
although the dual action (\ref{gtt}) possesses 
a noncommuting time variable in terms of 
the Moyal bracket (\ref{nc}) given by
$\ti{\ast},$
it is originated from the action (\ref{or}) 
whose time coordinate is commuting.
We wonder what would be the phase space structure if we
treat time coordinate as commuting in the action
(\ref{od}) written in components as
\beqa
S_d & = & -g^2\int 
d^4x\ 
[\fr{1}{2}F_{0i}F_{0i}-\fr{1}{4}F_{ij}F_{ij}
-\fr{1}{2}\ti{\te}^{0i}F_{i0}F_{0j}F_{0j}
-\ti{\te}^{0i}F_{ij}F_{jk}F_{k0} \nonumber \\
& &+\fr{1}{4}\ti{\te}^{0i}F_{i0}F_{jk}F_{kj}]. \lb{dao}
\eeqa
Definition of the spatial components of momentum
\beqa
P_{Di}=\fr{\kd 
\ti{S}}{\kd 
(\del_0 A_{Di})}& = &
g^2[F_{D0i}-\fr{1}{2}\ti{\te}^{0i}F_{D0j}F_{D0j}
-\ti{\te}^{0j}F_{Dj0}F_{D0i}+\ti{\te}^{0k}F_{Dkj}F_{Dji} \nonumber \\
& & +\fr{1}{4}\ti{\te}^{0i}F_{Djk}F_{Dkj} ] \lb{ao}
\eeqa
can be solved to find $\del_0A_{Di}$.
They lead 
to the same Hamiltonian 
 (\ref{sla})
which was obtained using the action (\ref{as}).
Moreover, there are the same constraints (\ref{chd}).

We conclude that at the first order
in $\ti{\te}$ 
whatever the method used 
we obtain the same Hamiltonian (\ref{sla}) 
and the constraints (\ref{chd}). 
However, the method of obtaining Hamiltonian 
from the shifted action  (\ref{as})
seems easier: 
When the higher orders in $\ti{\te}$  are considered
the unique change will be in the constraint (\ref{con4}), the
other constraints 
(\ref{con0})--(\ref{con2}), (\ref{con3})
will remain intact. Thus, finding  Hamiltonian
of the dual theory
is reduced to find  solution of a constraint.

\section{BPS States of  Non-commutative D3-brane}   

In the zero slope limit, $\al^\prime \rightarrow 0,$ and considering 
slowly varying fields, 
noncommutative
DBI action can be approximated as
 noncommutative gauge theory (\ref{sla}),
up to constant terms\cite{sw}.
Noncommutative D3--brane worldvolume action can be
extracted  from 10 dimensional noncommutative
gauge theory in the static gauge. The first
three spatial coordinates
are taken equal to brane worldvolume coordinates and 
the rest of the coordinates are regarded
as scalar fields on the brane.
We consider only one  scalar field.
D3--brane worldvolume 
Hamiltonian density resulting
from (\ref{sla}) when 
$\ti{\te}^{0i}\neq 0,\ \ti{\te}^{ij}= 0,$  is
\beqa
H&=&\fr{1}{2}{P_i}^2+\fr{1}{4}F_{ij}^2-\fr{1}{2}\ti{\te}^{0i}P_iP_j^2
-\fr{1}{4}\ti{\te}^{0i}P_iF_{jk}^2+\ti{\te}^{0i}F_{ij}F_{jk}P^k \nonumber \\
&& +\fr{1}{2}{\pi}^2+\fr{1}{2}(\del_i\phi)^2-\fr{1}{2}\ti{\te}^{0i}P_i{\pi}^2 
+\ti{\te}^{0i}\pi F_{ij}\del_j\phi \nonumber \\
&& -\ti{\te}^{0i}P_j\del_i\phi \del_j\phi +\fr{1}{2}
\ti{\te}^{0i}P_i{(\del_j\phi)}^2 \lb{hh} .
\eeqa
The scalar field and 
the corresponding
canonical momentum 
denoted as $\phi$ and
$\pi.$  Moreover,
we renamed the dual variables $F_D,\ P_D$ 
as $F,\ P.$
We choose $\pi=0$
to deal with the static case.

To discuss bounds on the
value of 
the Hamiltonian  we would like to write (\ref{hh})  
as
\be
\lb{hy}
H=\fr{1}{2}\widehat{P}_i^2+\fr{1}{2}\widehat{B}_i^2
+\fr{1}{2}(\widehat{\del_i\phi})^2 ,
\ee
with the restrictions
\[
\widehat{P}_i|_{P=0}=0,\ 
\widehat{B}_i|_{F=0}=0,\ 
\widehat{\del_i\phi}|_{\phi=0}=0.
\]
These are fulfilled by
\be
\lb{ht1}
\widehat{P}_i=P_i
-a_1\ti{\te}^{oi}P_j^2-
a_2\ti{\te}^{oj}P_j P_i,
\ee
\be
\lb{ht2}
\widehat{B}_i=\fr{1}{2}\epsilon_{ijk}(F_{jk}-
\fr{1}{2}\ti{\te}^{0l}P_l F_{jk}+b_1
\ti{\te}^{0l}P_k F_{jl}
+b_2
\ti{\te}^{0k}P_l F_{jl} ),
\ee
\be
\lb{ht3}
\widehat{\del_i \phi}=\del_i \phi
+\fr{1}{2}\ti{\te}^{0j}P_j \del_i \phi
-c_1\ti{\te}^{0i}\del_j \phi
 P_j-c_2\ti{\te}^{0j}\del_j \phi P_i ,
\ee
where  $a_{1,2}\ b_{1,2}\ c_{1,2}$ are constants which should satisfy
\be
\lb{cc}
a_1+a_2={1 \over 2},\ b_1+b_2=-2,\ c_1+c_2=1,
\ee
otherwise arbitrary. These do not correspond to the Seiberg--Witten map
(\ref{sw}). There
the fields of commutative and 
noncommutative gauge theories are mapped into each other
by changing  the gauge group from the ordinary $U(1)$
to noncommutative one such that (\ref{swc})
is satisfied.
In our case  gauge group is always $U(1).$
Although we write the Hamiltonian (\ref{hy}) in terms 
of $\ti{\te}^{0i}$
dependent  (noncommutative) fields  still there is the constraint 
\be
\lb{dp0}
\del_iP_i=0,
\ee
indicating $U(1)$ gauge group.
Seiberg--Witten map 
in phase space is studied in \cite{om}--\cite{ak}.

Now, in terms of an arbitrary angle $\alpha$ 
the Hamiltonian density (\ref{hh}) can be put into the form
\beqa
H & = & \fr{1}{2}( \widehat{P}_i-\sin \al\ \widehat{\del_i\phi})^2 
+\fr{1}{2}( \widehat{B}_i-\cos \al\ \widehat{\del_i\phi})^2 \nonumber \\ 
&& +\sin \al\  \widehat{P}_i\widehat{\del_i\phi} 
+\cos \al\  \widehat{B}_i\widehat{\del_i\phi} .
\eeqa
Thus, we can write a bound on the total energy $E$
relative to the worldvolume vacuum of noncomutative D3--brane
as
\be
\lb{bb}
E\geq \sqrt{{\ti{Z}_{el}}^2+{\ti{Z}_{mag}}^2},
\ee
where, we introduced
\beqa
\ti{Z}_{el} = \int_{D3} d^3x\ \widehat{P}_i\widehat{\del_i\phi}, \\
\ti{Z}_{mag} =  \int_{D3} d^3x\ \widehat{B}_i
\widehat{\del_i\phi} .
\eeqa
In the commutative case 
$\ti{Z}_{el}$ and $\ti{Z}_{mag}$ become topological 
charges due to the Gauss law
and the Bianchi identity:
$
\lb{gb}
\del_iP_i=0,\ 
\del_iB_i=0.$
In the commuting case (\ref{bb}) 
is known as  BPS bound\cite{gi},\cite{ggt}. However, in our case 
we do not have integrability conditions for 
$ \widehat{P}_i,\ \widehat{B}_i.$ 
Nevertheless, it will be shown that
$\ti{Z}_{el},\ \ti{Z}_{mag}$ can be 
topological charges for some specific configurations:

The bound (\ref{bb}) is saturated for 
\be
\lb{ncs1}
\widehat{P}_i=\widehat{\del_i\phi},\
\widehat{B}_i=0,\ \sin\al =1.
\ee
This can be accomplished 
at the 
first order in $\ti{\te}^{0i}$,
when
\be
\lb{bi}
F_{ij}=0,\
P_i= \del_i\phi ,
\ee
if  we fix the parameters as
\be
\lb{a}
a_1=c_1,\ a_2=c_2-{1\over 2},
\ee
which are consistent with (\ref{cc}).
Because of the constraint (\ref{dp0}), $\phi$ should satisfy
\be
\lb{le}
 \del_i^2\phi=0.
\ee
For this configuration $\ti{Z}_{mag}$
vanishes: $\ti{Z}^{(1)}_{mag}  =  0,$
and $\ti{Z}_{el}$ reads
\be
\ti{Z}^{(1)}_{el}  = 
\int_{D3} d^3x \del_i(\phi \del_i \phi )
- \int_{D3} d^3x\ \ti{\te}^{0i}\del_i \phi (\del\phi)^2.
\ee
For the commutative case isolated singularities of $\phi$
satisfying these conditions are called BIon\cite{gi}. The simplest choice
satisfying (\ref{le}) is\cite{ggt}
\be
\lb{obi}
\phi (r)={e \over 4\pi r} ,
\ee
where  $r$ is the radial
variable. In general we cannot write $\ti{\te}^{0i}$
dependent part as a surface integral.
However, this choice of harmonic function (\ref{obi}) 
renders it possible. Indeed,
we can write $\ti{Z}^{(1)}_{el}$ as an 
integral over a 
sphere of radius $\ep$ about the origin
and find
\be
\ti{Z}^{(1)}_{el} =
(e - {\ti{\te}e^2\over 20 \pi \ep^4} )
\lim_{\ep \rightarrow 0}\phi (\ep),
\ee
where $\ti{\te}\equiv \sqrt{\ti{\te}^{0i}\ti{\te}^{0i}}.$

Observe that the usual
 BIon solution (\ref{obi}) leads to a solution for the
noncommutative case (\ref{ncs1}).
This is similar to the fact that
linearized and 
full DBI actions lead to the same BIon 
solution with the same energy\cite{cm}. 
Here solutions are the same but energies differ.

When one sets $P_i=0$ the terms depending on the  noncommutativity parameter 
$\ti{\te}^{0i}$ disappear. This is what we expected: Noncommutativity is
only between time and space coordinates, not between 
spatial coordinates. Thus, when momenta vanish noncommutativity should cease to exist.
For 
$P_i=0,$
the bound (\ref{bb}) is saturated for
\be
\lb{hjk}
{1 \over 2} \ep_{ijk}F_{jk}= \del_i\phi ,\ \cos \al =1
\ee
where as before $\phi$ should satisfy (\ref{le}).
For this commuting configuration $\ti{Z}_{el}$ and 
$\ti{Z}_{mag}$ are given as
 $\ti{Z}^{(2)}_{el}  =  0, $ and
\be
\ti{Z}^{(2)}_{mag}  =  \int_{D3} d^3x \del_i(\phi \del_i \phi ) .
\ee
To satisfy (\ref{hjk}) and (\ref{le}) consider 
a magnetic charge at the origin
\be
\phi (r)={g \over 4\pi r} .
\ee
Let the  integral be over a 
sphere of radius $\ep$ about the origin
which yields
\be
\ti{Z}^{(2)}_{mag} = g\lim_{\ep \rightarrow 0}\phi (\ep).
\ee

There is another configuration 
\be
\lb{sol}
\widehat{P}_i=\sin \al\ \widehat{\del_i\phi} ,\ 
\widehat{B}_i=\cos\al\ \widehat{\del_i \phi},
\ee
which saturates the bound (\ref{bb}).
The constant angle $\al$ is defined as
\[
\tan\al={\ti{Z}_{el} \over \ti{Z}_{mag}}.
\]
This can be realized if the commuting variables
are fixed as
\be
P_i=\sin \al\ \del_i\phi ,\ 
{1 \over 2} \ep_{ijk}F_{jk}= \cos\al\ \del_i \phi  
\ee
and the free parameters 
in (\ref{ht1})--(\ref{ht3}) satisfy
(\ref{a}) and 
\be
c_1=b_1/2,\ c_2=1-b_1/2 .
\ee
These 
 are consistent with (\ref{cc}).
Thus, in the hatted quantities (\ref{ht1})--(\ref{ht3})
now, there is only one free constant parameter.
For this configuration  $\ti{Z}_{el}$ and $\ti{Z}_{mag}$ are given by
\beqa
\ti{Z}^{(3)}_{el} & = &
\int_{D3} d^3x \sin \al\ \del_i(\phi \del_i \phi )
- \int_{D3} d^3x\ \ti{\te}^{0i}\sin^2 \al\ \del_i \phi (\del\phi)^2,\lb{ze3}\\
\ti{Z}^{(3)}_{mag} & = & 
\int_{D3} d^3x \cos \al\ \del_i(\phi \del_i \phi ), 
- \int_{D3} d^3x\ \ti{\te}^{0i}\cos^2 \al\ \del_i \phi (\del\phi)^2. \lb{zm3}
\eeqa
Similar to the other configurations,
$\phi $ should satisfy (\ref{le}) and
we consider the simplest choice 
\be
\lb{dy}
\phi (r)={g \over 4\pi \cos \al\   r} .
\ee
For this choice of the harmonic function (\ref{dy})
the integrals in (\ref{ze3}) and (\ref{zm3}) can be performed   
over a sphere of radius $\ep$ about the origin.
Therefore, the energy can be calculated as
\be
E=\left[ 
(e - {\ti{\te}e^2\over 20\pi \ep^4} )^2 +
(g - {\ti{\te}g^2\over  20\pi \ep^4} )^2
\right]^{1/2}
\lim_{\ep \rightarrow 0}\phi (\ep),
\ee
where $e / g=\tan \al$. Similar to the above mentioned 
configurations ordinary
D3--brane dyon solution
(\ref{dy}), provide a solution of the
noncommutative condition
(\ref{sol}).

\section{Discussions}

The results which we obtained are valid  at the
first order in the noncommutativity parameter $\te.$ 
In principle contributions  at higher orders 
in $\te$ can be calculated. Obviously, one of the methods is
to solve $\del_0A_D$ in terms of $P_D,\ F_D$
from the generalization of (\ref{ao}).
However, it is highly non--linear.
On the other hand using the shifted 
action as it is illustrated here 
seems more manageable.
We are encouraged from the fact that one should only 
solve a constraint similar to (\ref{con4}). The other constraints 
(\ref{con0})--(\ref{con2}),(\ref{con3})
remain intact.

Noncommuting D3--brane formulation
which we deal with is somehow
different from the one considered in \cite{mat}, 
\cite{ha}--\cite{hh}. There,
gauge group is noncommutative $U(1)$, in our case although 
Hamiltonian depends on the  noncommutativity parameter $\te,$ gauge
group is still $U(1).$ 
This seems to be the basic reason that the  
BPS solutions
of ordinary case\cite{gi}--\cite{ggt} 
provide  solutions of the noncommutative
case as it happens between linearized and full
DBI action\cite{cm}.

\vspace{2cm}

\noindent
{\bf Acknowledgment}

\vspace{.5cm}

\noindent
We thank I. H. Duru  for his kind helps.

\end{document}